\documentclass[preprint,12pt,3p]{elsarticle}

\usepackage{matlab-prettifier}
\usepackage{amsmath}
\usepackage{xfrac}
\usepackage{graphicx}
\usepackage{parskip}
\usepackage{hyperref}
\usepackage{longtable}
\usepackage{adjustbox}
\usepackage{setspace}
\hypersetup{ 
backref=true, 
bookmarks=true, 
unicode=false, 
pdftoolbar=true, 
pdfmenubar=true, 
pdffitwindow=false, 
pdfstartview={FitH}, 
pdftitle={}, 
pdfauthor={}, 
pdfsubject={}, 
pdfkeywords={}{}, 
pdfnewwindow=true, 
colorlinks=true, 
linkcolor=BlueViolet, 
citecolor=maroon, 
urlcolor=blue,
}
\usepackage{comment}
\usepackage{subfig}
\usepackage{caption}

\usepackage{amssymb}
\usepackage{gensymb}

\usepackage{natbib}
\biboptions{comma,round}
\setcitestyle{authoryear}

\begin{document}
\begin{frontmatter}

\title{The climate niche of \textit{Homo Sapiens}}

\author[label1,label2,label3,label4,label5,label6,label7]{Richard S.J. Tol}
\address[label1]{Department of Economics, Jubilee Building, University of Sussex, Falmer, BN1 9SL, United Kingdom; r.tol@sussex.ac.uk}
\address[label2]{Institute for Environmental Studies, Vrije Universiteit, Amsterdam, The Netherlands}
\address[label3]{Department of Spatial Economics, Vrije Universiteit, Amsterdam, The Netherlands}
\address[label4]{Tinbergen Institute, Amsterdam, The Netherlands}
\address[label5]{CESifo, Munich, Germany}
\address[label6]{Payne Institute for Public Policy, Colorado School of Mines, Golden, CO, USA}
\address[label7]{College of Business, Abu Dhabi University, UAE}

\begin{abstract}
I propose the Dominicy-Hill-Worton estimator to estimate the current climate niche of \textit{Homo Sapiens} and our croplands. I use this to extrapolate the degree of unprecedentedness of future climates. Worton's peeled hull is a non-parametric, N-dimensional generalization of order statistics. Dominicy and colleagues show that Hill's estimator of the tail-index can be applied to any homogeneous function of multivariate order statistics. I apply the Dominicy-Hill estimator to transects through Worton's peels. I find a thick tail for low temperatures and a thin tail for high ones. That is, warming is more worrying than cooling. Similarly, wettening is more worrying than drying. Furthermore, temperature changes are more important than changes in precipitation. The results are not affected by income, population density, or time. I replace the Hill estimator by the QQ one and correct it for top-censoring. The qualitative results are unaffected.\\
\\
\textit{Keywords}: tail-index, convex hull, climate\\ \\
\textit{JEL codes}: C46, N50, O44, Q54
\end{abstract}

\end{frontmatter}

\newpage\section{Introduction}
\label{sc:intro}
A key concern about greenhouse gas emissions is that future climate will be \emph{unprecedented} in the evolutionary history of humankind. It is one of the three justifications for the 2\celsius{} target of the Paris Agreement \citep[][see also \citet{Tol2007EP}]{WBGU1995} and has raised worries about the safety of humankind \citep{Rockstrom2009}, inspiring such organizations as \textit{Extinction Rebellion}. Unprecedented is not synonymous with detrimental, however \citep{Hume1739, Moore1903}. Furthermore, as the \emph{physical} evolution of \textit{Homo Sapiens} is slow because of our slow maturation and low birth numbers, behavioural adaptation is well-developed. Humans are very adaptive and actively shape their environment to suit their needs, traits shared, to a lesser extent, by other long-lived animals. In this paper, I develop a method to assess how a species would fare outside its observed climate niche, and apply this to humans and their sustenance.

\begin{figure}[h]
\centerline{\includegraphics[width=\textwidth]{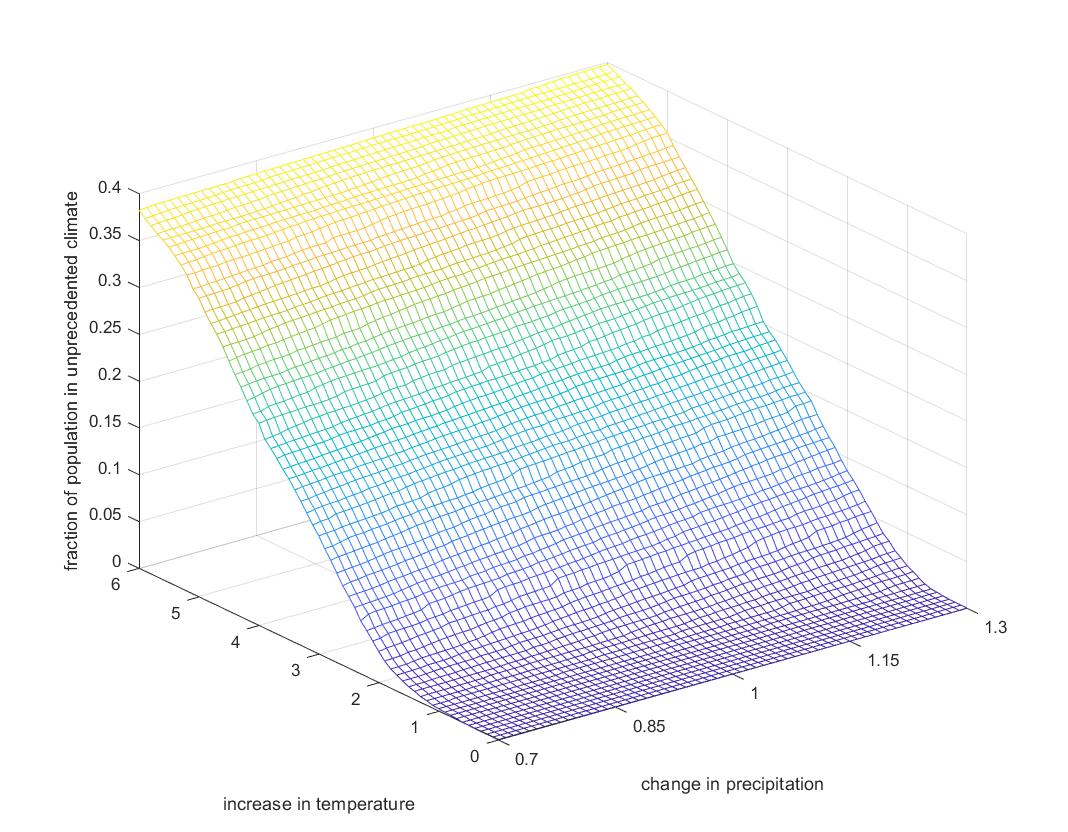}}
\caption{The fraction of the human population that would be exposed to an unprecedented climate as a function of absolute warming and relative drying or wettening.}
\label{fig:extrapolate}
\end{figure}

Figure \ref{fig:extrapolate} illustrates this concern. It takes the current distribution of humans in climate space\textemdash here defined as the annual average temperature and the annual total precipitation\textemdash and computes the fraction of humanity that would experience a climate where no human has lived before should temperature increase and rainfall change. Figure \ref{fig:extrapolate} shows two things. First, precipitation does not really matter in this regard. Humans live in such a wide variety of rain and snow climates that a 30\% increase or a 30\% decrease would push few people into unprecedented territory. The opposite is true for temperature. Many people live close to the observed maximum temperature so that 6\celsius{} of warming would push 40\% of people into unchartered climate territory. Note also that 1\celsius{} of warming\textemdash roughly 2\celsius{} above preindustrial\textemdash would have a minimal effect.

\citet{Xu2020} study the same topic using roughly the same data. They argue that human life is fragile to climate change. Their estimate of the human climate envelope is unfortunately biased. They removed the extreme 1\% of the data but did not correct their estimates for censoring. Relying on a Gaussian mixture, they assume that tails are thin. Unprecedented climates are therefore harsh \emph{by assumption}. This is strengthened by Xu's assumption that the human climate niche is the 90\% confidence interval of their population model. In other words, they assume that 10\% of the (predicted) world population live in unlivable conditions, an inconsistent assumption. These are not zombies. In contrast, I do not censor the data, I assume that people are alive regardless of where they are recorded to live, and I estimate rather than assume the thickness of the tail. Like \citet{Xu2020}, I find that a warmer and wetter climate would pose problems for humanity\textemdash but my numbers of people are threat are much smaller.

The paper proceeds as follows. Section \ref{sc:methods and data} discusses methods and data. Section \ref{sc:results} shows results for people and crops. Section \ref{sc:implications} presents the implications of future climate change. Section \ref{sc:conclude} concludes.

\section{Methods and data}
\label{sc:methods and data}
\subsection{Methods}
\label{sc:methods}
\begin{figure}[h]
\centerline{\includegraphics[width=\textwidth]{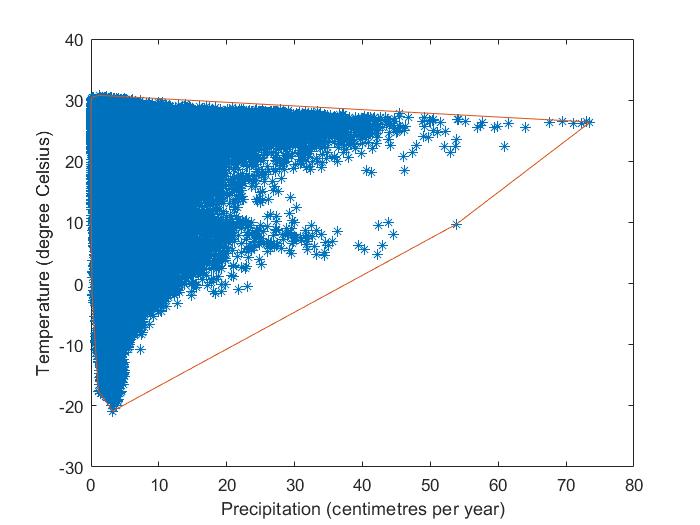}}
\caption{Human occupation of temperature-precipitation space, 2000, and its convex hull.}
\label{fig:convexhull}
\end{figure}

How to estimate a niche? You could define a niche as a stochastic frontier. Originally, stochastic frontier analysis estimates the \emph{maximum} possible production, a frontier that cannot be exceeded. Flipping the sign, you can estimate a minimum too. Taking the distance from a central point, you can estimate a perimeter, an area you cannot go outside. The stochastic frontier is parametric, however. If it is approximated by a second-order polynomial in explanatory variables, the niche would be an ellipsoid around the central point. Figure \ref{fig:convexhull} shows the irregular shape of human occupation of climate space in the year 2000, here limited to average temperature and total precipitation. An ellipse badly approximates these observations.

Alternatively, you could construct a multivariate kernel density. However, kernel densities are excellent for describing data but unsuitable for extrapolation. Furthermore, there are parts of climate space that are not occupied by humans but surrounded by occupied areas. Figure \ref{fig:convexhull} shows that no human lives at 12\textdegree C and 25 cm/year\textemdash but humans live in colder or hotter places that are just as wet, and in wetter and dryer places that are just as warm. A kernel density would suggest that it is impossible to live at 12\textdegree C and 25 cm/year\textemdash although in fact no one lives there is, currently, no such climate on Earth.

Therefore, I use the convex hull, a linear combination of observations that envelops all observations. This is also shown in Figure \ref{fig:convexhull}. The convex hull is a multi-dimensional generalization of the uni-dimensional minimum and maximum. In one dimension, all observations lie between the minimum and maximum. In two (or more) dimensions, all observations lie within the convex hull.

\begin{figure}[h]
\centerline{\includegraphics[width=\textwidth]{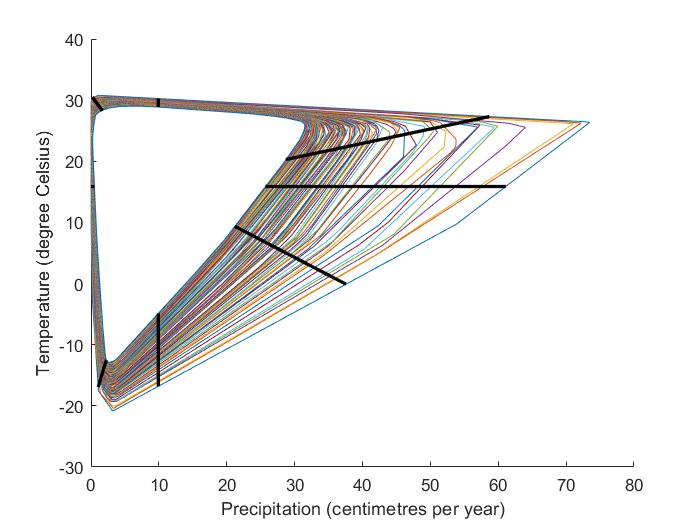}}
\caption{The peeled hull of the human occupation of climate space, 2000, and the studied order statistics of temperature and precipitation.}
\label{fig:peeledhull}
\end{figure}

Like a kernel density, a convex hull is a purely descriptive device with little scope for extrapolation. A convex hull is an extremum. If the data were a sample, we could derive its distribution and use that to extrapolate. However, the data used here are the population, rather than a sample. \citet{Worton1995} suggested a more promising route: The peeled hull. The convex hull is a subset of the set of observations. Removing the convex hull from the set of observations gives a reduced sample, with its own convex hull. You can continue to ``peel'' the observations until only, say, 95\% of the sample is left. That is, just as the convex hull generalizes the minimum and maximum, the \emph{peeled} hull generalizes the order statistics. Figure \ref{fig:peeledhull} shows the peeled hulls; the smallest includes 95\% of the human population.

\citet{Dominicy2017} show that the \citet{Hill1975} estimator of the tail-index can be applied to any \emph{norm} of the multivariate order statistics. \citet{Dominicy2020} generalize this to any \emph{homogeneous function}\textemdash Hill's statistic, divided by the degree of homogeneity, is a consistent estimator of the tail index. \citet{Dominicy2020} repurpose the algorithm by \citet{Rousseeuw1999} to define the order statistics. However, that algorithm is parametric; the order statistics are ellipsoids. I therefore prefer the peeled hull of \citet{Worton1995}.

I apply the proposed Dominicy-Worton-Hill estimator to the area of the convex hull and the length of its perimeter, and to the eight trajectories shown in Figure \ref{fig:peeledhull}. The trajectories together span all possible climate change: warmer, colder, wetter, drier, and any combination. The trajectories are defined as the intersection of the peeled hulls with straight lines from the population-weighted average temperature and precipitation to their maximum and minimum.

The distance between two points on a trajectory is homogeneous of degree one. The perimeter length is the sum of such distances and so homogeneous of degree one too. The area is homogeneous of degree two, which is easily seen by dividing the area into triangles.

\citet{Hill1975} proposed the maximum likelihood estimator of the tail-index, valid if the tail of the distribution is exactly Pareto. A quantile-quantile (QQ) estimator is preferred if the distribution is only approximately Pareto. For robustness, I here use the QQ estimator proposed by \citet{Brito2003}: The geometric average of the coefficients from (1) regressing the natural logarithm of size on the natural logarithm of its rank and (2) regressing log rank on log size.

The former estimator was proposed by \citet{Kratz1996}. It naturally generalizes to the case of censored data \citet{Tobin1958, Goldberger1964}, as does the inverse estimator. This Tobit model is relevant as people have occupied the hottest and driest parts of the planet\textemdash and may have moved to hotter and drier places if available.

\subsection{Data}
\label{sc:data}
The two main data sets are CRU TS and HYDE. HYDE 3.2 \citep{Klein2010} reports human population counts and densities for the whole world, at a $5' \times 5'$ grid, for the last 12,000 years, in time steps of 1,000 (100) years before (after) the birth of Christ. In 10,000 BC, 228,563 grid cells were occupied by humans, increasing to 277,987 grid cells in 2000. HYDE also reports croplands and urbanization.

CRU TS 4.05 \citep{Harris2020} reports average monthly temperature and total monthly precipitation at a $30' \times 30'$ grid for 1901 to 2020. I compute the annual average for the period 1991-2020, as climate is typically defined as the thirty-year average weather.

Assuming a homogeneous climate within the CRU grid cells, I overlay CRU and HYDE data and then aggregate the population number and grid cell area to every unique combination of temperature and precipitation; average population density by climate cell follows readily. This is the main database: human population in climate space. I follow the same procedure for cropland.

I also consider per capita income, from the Global Data Lab (GDL)  \citep{Smits2019}. They report per capita income, in Geary-Khamis dollars reflecting local purchasing power, for subnational administrative units. GDL has a crude spatial resolution compared to \citet{Kummu2018} and \citet{Nordhaus2006}. However, GDL reports \emph{income} while Kummu and Nordhaus report \emph{output} as measured by value added. The difference between income and output is particularly large in inner cities where many work but few live. \citet{Harris2020} removed the urban heat island effect \citep{Estrada2017} from their data. Output and income also differ in the oil fields of Siberia and the Arabian Desert, which would distort any analysis of the impact of climate on economic \emph{output} per person. I therefore use GDL data. Assuming a uniform income distribution within provinces and states, I overlay GDL and HYDE data and the result with CRU data to find human population in climate space by income. 

\section{Results}
\label{sc:results}

\subsection{People}
Table \ref{tab:area} shows the estimated tail-indices for area and perimeter. The estimates vary between 6.4 and 7.4, indicating a thin tail: The first 6 moments exist, but the 8th moment does not. Table \ref{tab:traject} shows a more nuanced picture: Tail-indices are estimated separately for temperature and precipitation, and separately for left and right tail. Table \ref{tab:traject} reveals a thin tail for temperature, and a thinner one for hot conditions than for cold ones. The tail for precipitation is less thin for wet conditions and thick for dry conditions.

A thick tail means that there is a reasonable chance of observing something much larger than was observed before. Thick tails are usually a source of concern. A thick tail for stock market returns means the market can drop far fast. A thick tail for rainfall means the next flood may be much worse than anything seen before. In this application, however, thick tails are a good sign. Humans can reasonably be expected to live in areas drier than observed. On the other hand, thin tails are a reason for concern. Human population drops rapidly in hotter conditions. Higher temperatures still would pose a real problem.

\begin{figure}[h]
\centerline{\includegraphics[width=\textwidth]{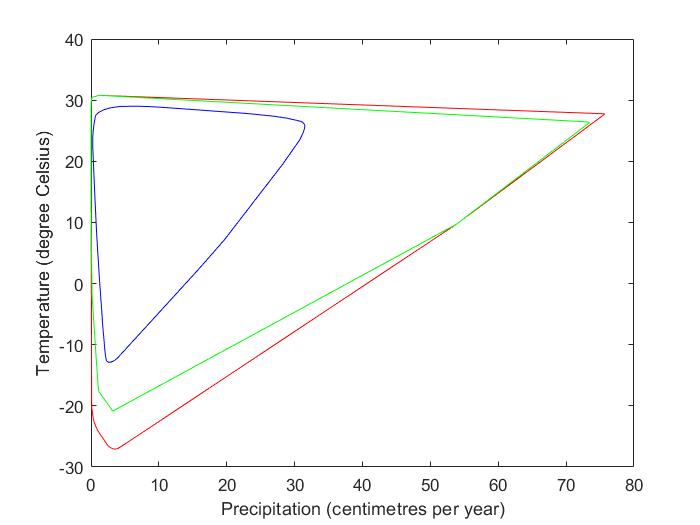}}
\caption{The convex hull of the human occupation of climate space, 2000, the convex hull that contains 95\% of all humans, and the convex hull of temperature and precipitation.}
\label{fig:climatehull}
\end{figure}

This conclusion is both illustrated and called into question by Figure \ref{fig:climatehull}, which shows three convex hulls. Comparing the hull that contains all of humanity to the one that contains 95\% of people, we see that the order statistics span a large distance over wet conditions, a smaller distance over cold conditions, and hardly any distance over hot conditions. This explains the estimated relative thinness of the tails. The third convex hull in Figure \ref{fig:climatehull} is for the observed temperature and rainfall on the six inhabited continents. Humans have occupied almost the entire available climate space, except for the very cold parts of the planet and its very wet and hot bits. This contradicts the thin tail for heat. We could have lived, indeed may have lived, in hotter places if these were there.

\begin{figure}[h]
\centerline{\includegraphics[width=0.8\textwidth]{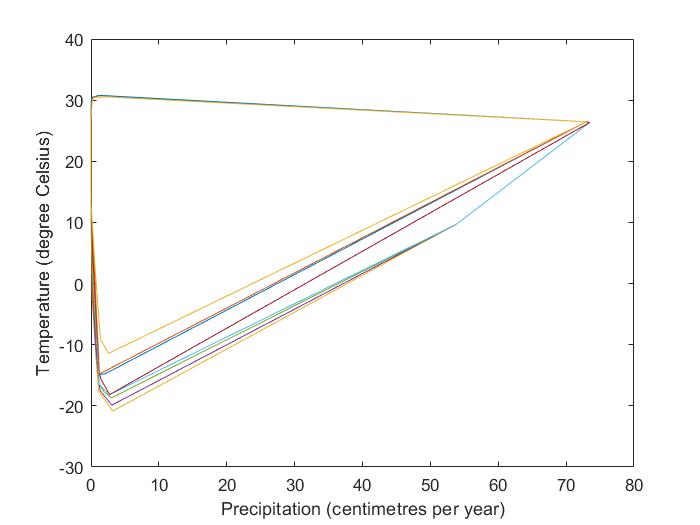}}
\centerline{\includegraphics[width=0.8\textwidth]{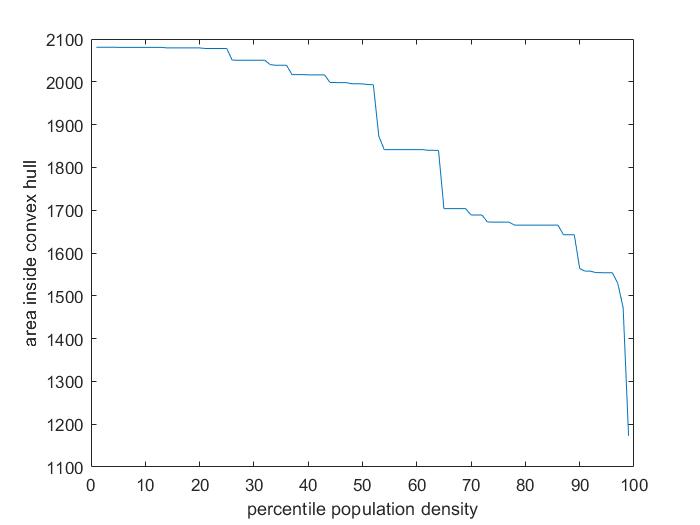}}
\caption{The convex hull of the human occupation of climate space in 2000 for all population densities and its deciles (top panel) and the area of hull for the percentiles of the distribution of population density (bottom panel).}
\label{fig:popdens}
\end{figure}

The latter interpretation is supported by Figure \ref{fig:popdens}. The upper panel shows the convex hull for the whole population, the population in areas where the population density is at the 90\%ile or lower, and so on all the way down to the 10\%ile. The convex hull is curtailed at the cold and wet end. The lower panel shows the area of the convex hull, removing areas in steps of 1\%ile. The convex hull starts to really change only when we exclude everywhere below the median population density\textemdash and even restricting attention to the most densely populated cities does not halve the area of the convex hull.

Table \ref{tab:tobit} supports the opposite interpretation. It shows estimates of the tail-index with and without correcting for censoring the temperature and precipitation data. Although estimates are different, these differences are not statistically significant. The qualitative conclusion remains: The tail of the temperature distribution is thin, also if we take into account that humans live in the hottest places on Earth and therefore could not have lived in even hotter places.

\begin{figure}[h]
\centerline{\includegraphics[width=0.8\textwidth]{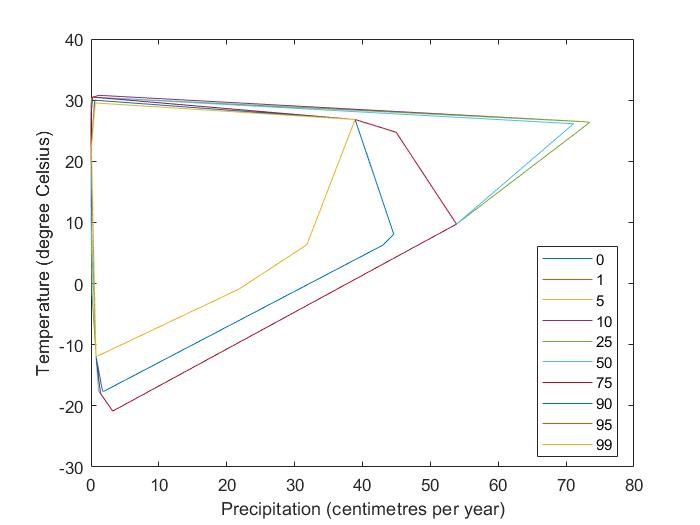}}
\centerline{\includegraphics[width=0.8\textwidth]{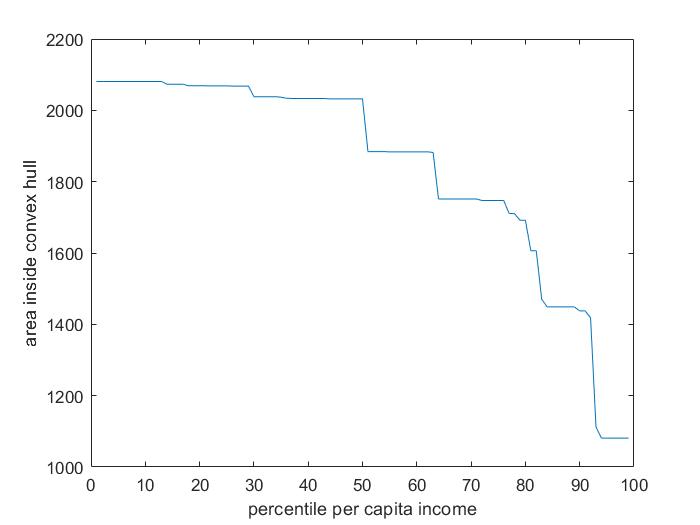}}
\caption{The convex hull of the human occupation of climate space in 2000 for all levels of per capita income and its deciles (top panel) and the area of hull for the percentiles of the distribution of per capita income (bottom panel).}
\label{fig:income}
\end{figure}

Figure \ref{fig:income} is similar to Figure \ref{fig:popdens} but sample restrictions are based on per capita income rather than population density. As above, the convex hull changes shape at the cold and wet extremes, but not so much at the hot and dry extremes. The convex hull really changes only when we restrict attention to the 20\% richest people.

\begin{figure}[h]
\centerline{\includegraphics[width=\textwidth]{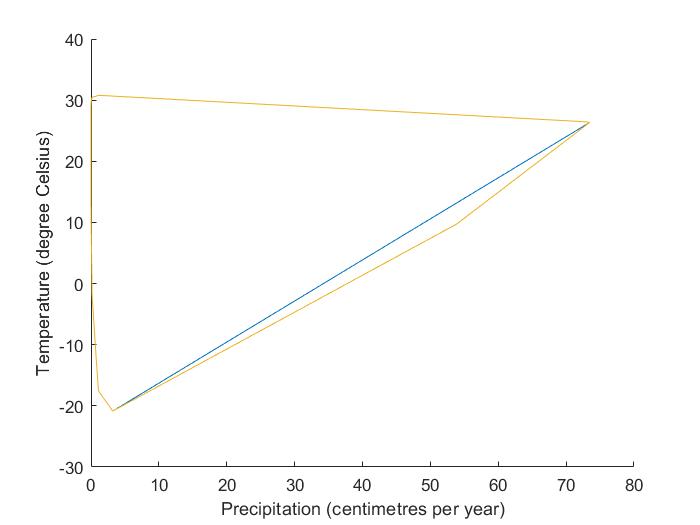}}
\caption{The convex hull of the human occupation of climate space before and after 1200 AD.}
\label{fig:historyhull}
\end{figure}

Figure \ref{fig:historyhull} shows the convex hull for all years in the HYDE data, but using modern climate. The convex hulls are the same for the years 10,000 BC to 1000 AD and for the years 1100 AD to 2000 AD. The settlement of Bergen, Norway, in the 11th century makes the difference: Bergen is not particularly cold but it is exceptionally wet. Figure \ref{fig:income} shows that income and \emph{access} to technologies to weather extreme climates do not really affect where people live. Figure \ref{fig:historyhull} shows that technology itself has been sufficiently mature for 12,000 years (or more) to sustain people in very hot, very cold, very dry, and very wet places.

\subsection{Crops}
Tables \ref{tab:croparea} and \ref{tab:croptraject} show the estimated tail-indices for area and perimeter, and for the trajectories across the peeled hulls for croplands in the year 2000. The results are very similar to the ones for human occupation, except that tails are somewhat thinner for crops than for humans. Cooling is less of a concern than warming, drying is less worrying than wettening, and temperature is more important than precipitation.

\begin{figure}[h]
\centerline{\includegraphics[width=\textwidth]{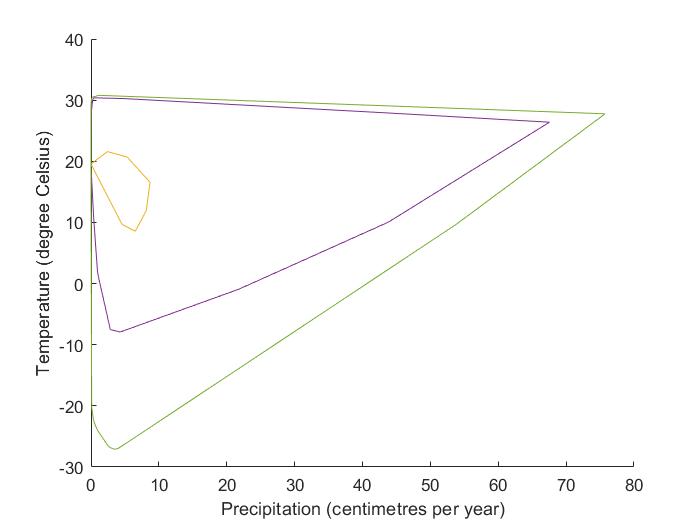}}
\caption{The convex hull of cropland in climate space in 5000 BC and 2000 AD, and the convex hull of temperature and precipitation.}
\label{fig:crophull}
\end{figure}

Figure \ref{fig:crophull} shows three convex hulls. The largest is the convex hull of all climates on Earth. The second-largest contains all croplands in the year 2000. A comparison with Figure \ref{fig:historyhull} shows that there are places with humans but no crops. This may well imply that there are climates where humans have tried and failed to grow crops. This is a worrying sign for climate change.

The smallest convex hull in Figure \ref{fig:crophull} contains all croplands in the year 5000 BC. It is small. While research on the impact of climate change on agriculture and studies of technological progress in agriculture have almost exclusively focused on the \emph{internal} margin\textemdash where crops are currently grown\textemdash Figure \ref{fig:crophull} reveals remarkable progress on the \emph{external} margin\textemdash developing varieties and methods to grow crops where they previously would not. While remarkable, the speed is slow, in the order of 1\celsius{} per millennium, much slower than the rate of warming expected for the 21st century.

\section{Implications}
\label{sc:implications}
\begin{figure}[h]
\centerline{\includegraphics[width=\textwidth]{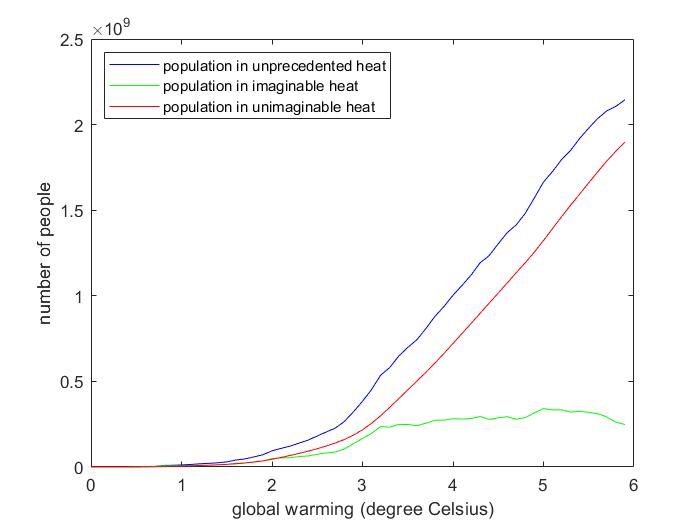}}
\caption{The number of people exposed to unprecedented heat, unprecedented but imaginable heat, and unprecedented and unimaginable heat.}
\label{fig:warming}
\end{figure}

Figure \ref{fig:extrapolate} shows that temperature is more important than precipitation in driving humankind out of our climate niche. The analysis therefore proceeds with temperature only. The top graph in Figure \ref{fig:warming} repeats the information from Figure \ref{fig:extrapolate} but in absolute numbers. Assuming no population growth, over two billion people would live in unprecedented heat should the world warm by 6\celsius. Note again how 1\celsius{} warming above today would have a minimal effect.

Unprecedented is a sharp divide; people either live in a certain climate or they do not. The statistical analysis above allows for a more gradual transition, a measure of the degree of unprecedentedness. The bottom graph applies the Pareto distribution estimated above; see Table \ref{tab:traject}. The tail-index indicates the speed of decline of the likelihood of finding people in a certain temperature, as the temperature rises. Normalizing this to unity at the convex hull\textemdash the maximum observed temperature is livable as shown by the people who live there\textemdash I then calculate the ``probability'' that people could live at higher temperatures. I refer to this as the number of people living in unprecedented but imaginable heat\textemdash where imagination is based on extrapolation from experienced circumstances. This number is zero by construction without climate change, but increases to an expected value of 300 million for 5\celsius{} of warming.

The difference between the two graphs is the number of people who would live in unprecedented and unimaginable heat. This number rises to slightly \emph{less} than 2 billion people for a global warming of 6\celsius.

\section{Discussion and conclusion}
\label{sc:conclude}
What climate is suitable for human habitation? I answer that question by mapping people and their crops on the temperature-precipitation space. I use Worton's peeled convex hull to find the climate niche of \textit{Homo Sapiens} and its order statistics, and the Dominicy-Hill estimator to determine the thickness of the tails of the distribution of humans in climate space. I find thin tails at hot end of the distribution, somewhat less thin tails at the cold end, thicker tails at the wet end, and fat tails at the dry end. As future climate is almost certainly hotter and probably wetter, this pattern is the opposite of what would be good for humans. Controlling for population density or average income does not affect the qualitative findings, which also appear to be stable over time and robust to censoring. The distribution of croplands shows the same results, with slightly thinner tails. Extrapolating to a warmer world, some 200 million people would be exposed to unprecedented heat, of which about half would experience unimaginable heat, if the world would warm by a further 2\celsius.

Although qualitatively similar to the results of \citet{Xu2020}, my quantitative results are very different. \citet{Xu2020} put 1-3 billion people outside the historical climate niche in the next fifty years, more than an order of magnitude more than the findings here. The reason is that \citet{Xu2020} assume that 10\% of the current population already live outside the climate niche. My starting point is that everyone alive is in livable conditions.

There are a number of caveats. Climate is measured by annual average temperature and annual total precipitation. Although many climate variables are correlated with these two variables, that correlation is not perfect and a richer set of variables may well lead to somewhat different conclusions. More importantly, there is no agency in the model and no microclimate. Humans migrate \citep{Barrios2006, Marchiori2012}, adjust their behaviour to the weather \citep{Adger2003, Auffhammer2014, Kousky2014, Zivin2014}, seek cooler places in their immediate environment \citep{Watanabe2016, Yang2017}, and change their environment \citep{Estrada2017}. These features partly determine where people live and are thus included in the statistical analysis. The extrapolation, however, implicitly assumes that future human behaviour is like current human behaviour.

These caveats notwithstanding, the results suggest that climate change poses an existential threat to many people. That these are measured in the hundreds of millions rather than billions as in previous studies, is scant comfort. The proposed method applied to other species, that are no generalists like \textit{Homo Sapiens}, is likely to show much larger impacts.

\bibliography{master}

\begin{table}[h]
    \centering
\begin{tabular}{l r r} \hline
     & Area & Perimeter \\ \hline
     Hill & 6.27 & 6.71 \\
     & (0.67) & (0.72) \\
     QQ & 7.38 & 7.22 \\
     & (0.08) & (0.02) \\ \hline
\end{tabular}
\caption{Estimates of the tail-index of the area and the perimeter length of the peeled convex hull for the human population in the year 2000.}
    \label{tab:area}
\end{table}

\begin{table}[h]
    \centering
\begin{tabular}{l r r r r r r r r} \hline
     & Hotter & Hotter & & Colder & Colder & Colder & & Hotter \\ 
     & & Wetter & Wetter & Wetter & & Dryer & Dryer & Dryer \\ \hline
     \multicolumn{9}{c}{Temperature} \\ 
     Hill & 628.99 & 139.99 & & 40.67 & 34.16 & 89.71 & & 311.34 \\
     & (67.04) & (14.86) & & (4.33) & (3.64) & (9.56) & & (33.18) \\
     QQ & 802.37 & 144.75 & & 113.35 & 80.62 & 240.07 & & 500.51 \\
     & (0.00) & (0.00) & & (0.00) & (0.00) & (0.00) & & (0.00)\\
     \multicolumn{9}{c}{Precipitation} \\
     Hill & & 3.95 & 4.04 & 5.20 & & 1.86 & 0.00 & 0.65\\
     & & (0.42) & (0.43) & (0.55) & & (0.20) & (0.00) & (0.07)\\
     QQ & &  4.68 & 4.39 & 6.17 & & 6.06 & 1.05 & 2.11 \\
     & & (0.03) & (0.03) & (0.02) & & (0.02) & (0.14) & (0.07)\\ 
     \hline
\end{tabular}
\caption{Estimates of the tail-index of temperature and precipitation along the trajectories of Figure \ref{fig:peeledhull} for the human population in the year 2000.}
    \label{tab:traject}
\end{table}

\begin{table}[h]
    \centering
\begin{tabular}{l r r r r r r r r} \hline
	&	Hotter	&	Hotter	&		&	Colder	&	Colder	&	Colder	&		&	Hotter	\\
	&		&	Wetter	&	Wetter	&	Wetter	&		&	Dryer	&	Dryer	&	Dryer	\\ \hline
\multicolumn{9}{c}{Temperature}\\
OLS	&	828.98	&	147.13	&		&	142.51	&	93.59	&	272.57	&		&	529.16	\\
	&	(122.23)	&	(21.69)	&		&	(21.01)	&	(13.80)	&	(40.19)	&		&	(78.02)	\\
Tobit	&	783.58	&	138.56	&		&	142.13	&	93.39	&	272.08	&		&	490.87	\\
	&	(115.53)	&	(20.43)	&		&	(20.96)	&	(13.77)	&	(40.12)	&		&	(72.37)	\\
\multicolumn{9}{c}{Rainfall}\\
OLS	&		&	4.79	&	4.42	&	6.33	&		&	7.07	&	1.71	&	2.65	\\
	&		&	(0.71)	&	(0.65)	&	(0.93)	&		&	(1.04)	&	(0.25)	&	(0.39)	\\
Tobit	&		&	4.47	&	4.23	&	6.33	&		&	7.06	&	1.71	&	2.65	\\
	&		&	(0.66)	&	(0.62)	&	(0.93)	&		&	(1.04)	&	(0.25)	&	(0.39)	\\
     \hline
\end{tabular}
\caption{Estimates of the tail-index of temperature and precipitation along the trajectories of Figure \ref{fig:peeledhull} for the human population in the year 2000 with (Tobit) and without (OLS) correcting for censoring.}
    \label{tab:tobit}
\end{table}

\begin{table}[h]
    \centering
\begin{tabular}{l r r} \hline
     & Area & Perimeter \\ \hline
     Hill & 7.69 & 7.63 \\
     & (0.33) & (0.33) \\
     QQ & 9.90 & 10.15 \\
     & (0.02) & (0.01) \\ \hline
\end{tabular}
\caption{Estimates of the tail-index of the area and the perimeter length of the peeled convex hull for cropland in the year 2000.}
    \label{tab:croparea}
\end{table}

\begin{table}[h]
    \centering
\begin{tabular}{l r r r r r r r r} \hline
     & Hotter & Hotter & & Colder & Colder & Colder & & Hotter \\ 
     & & Wetter & Wetter & Wetter & & Dryer & Dryer & Dryer \\ \hline
     \multicolumn{9}{c}{Temperature} \\ 
     Hill & 1129.6 & 184.1 & & 54.7 & 44.0 & 100.4 & & 401.9 \\
     & (49.1) & (8.0) & & (2.4) & (1.9) & (4.4) & & (17.5) \\
     QQ & 1574.3 & 240.6 & & 207.5 & 128.5 & 383.6 & & 478.8 \\
     & (0.0) & (0.0) & & (0.0) & (0.0) & (0.0) & & (0.0)\\
     \multicolumn{9}{c}{Precipitation} \\
     Hill & & 4.93 & 5.34 & 6.70 & & 2.13 & 0.42 & 0.45\\
     & & (0.21) & (0.23) & (0.29) & & (0.09) & (0.02) & (0.02)\\
     QQ & &  7.18 & 6.87 & 8.34 & & 9.37 & 2.22 & 3.36 \\
     & & (0.01) & (0.01) & (0.01) & & (0.01) & (0.03) & (0.02)\\ 
     \hline
\end{tabular}
\caption{Estimates of the tail-index of temperature and precipitation along the trajectories of Figure \ref{fig:peeledhull} for cropland in the year 2000}.
    \label{tab:croptraject}
\end{table}

\end{document}